\documentclass[prd,eqsecnum,leqnum,floats]{revtex4}

\begin{document}

\title{Application of Lagrange mechanics for analysis of the light-like particle motion in
pseudo-Riemann space}
\author{ Wladimir Belayev \\
 \normalsize Center for Relativity
and Astrophysics, Saint-Petersburg, Russia \\e-mail:
wbelayev@yandex.ru }

 \begin{abstract}We consider variation of energy of
the light-like particle in the pseudo-Riemann space-time, find
Lagrangian, canonical momenta and forces. Equations of the
critical curve are obtained by the nonzero energy integral
variation in accordance with principles of the calculus of
variations in mechanics. This method is compared with the Fermat's and geodesics principles. Equations for energy and momentum of the particle transferred to the gravity field are defined. Equations of the critical curve are solved for
the metrics of Schwarzschild, FLRW model for the flat space and
Goedel. The gravitation mass of the photon is found in central
gravity field in the Newtonian limit.
\end{abstract}

\maketitle

\\

\section{Introduction}

One of postulates of general relativity is claim that in gravity field in the absence of other forces the word lines of the material particles and the light rays are geodesics. In differential geometry a geodesic line in case of not null path is defined as a curve, whose tangent vector is parallel propagated along itself [1]. The differential equations of geodesic can be found also by the variation method as a path of extremal length with the aid of the virtual displacements of coordinates $x^{i} $ on a small quantity $\omega ^{i} $. When we add variation to coordinate of the material particle, the time-like interval slow changes, though that leaves it time-like.

Finding of the differential equations of the null geodesic, corresponding to the light ray motion, by calculus of variations is described in [2]. In space-time with metrical coefficients $g_{ij} $ it is considered variation of the first integral of these equations
\begin{equation} \label{GrindEQ__1_}
\eta =g_{ij} \frac{dx^{i} }{d\mu } \frac{dx^{j} }{d\mu } ,
\end{equation}
where $\mu $ is affine parameter. Deriving variation for extremum
determination we must admit arbitrary small displacements of
coordinates. The variation of integral of $\eta $ expanded in
multiple Taylor series is written as

\[{ \delta }I=\int _{\mu _{0} }^{\mu _{1} } \left\{\sum _{n=1}^{\infty } \sum _{\beta _{1} +\ldots +\beta _{4} =n} \left[\frac{1}{\beta _{1} !\ldots \beta _{4} !} \frac{\partial ^{n} g_{ij} }{\partial ^{\beta _{1} } x^{1} \ldots \partial ^{\beta _{4} } x^{4} } \frac{{ d}x^{i} }{{ d}\mu } \frac{{ d}x^{j} }{{ d}\mu } (\omega ^{1} )^{\beta _{1} } \ldots (\omega ^{4} )^{\beta _{4} } \right]+\right. \]
\[+g_{ij} \left(2\frac{{ d}x^{i} }{{ d}\mu } \frac{{ d}\omega ^{j}
}{{ d}\mu } +\frac{{ d}\omega ^{i} }{{ d}\mu } \frac{{ d}\omega
^{j} }{{ d}\mu } \right)+ \]
\begin{equation} \label{GrindEQ__2_}
\left. +\sum _{n=1}^{\infty } \sum _{\beta _{1} +\ldots +\beta
_{4} =n} \left[\frac{1}{\beta _{1} !\ldots \beta _{4} !}
\frac{\partial ^{n} g_{ij} }{\partial ^{\beta _{1} } x^{1} \ldots
\partial ^{\beta _{4} } x^{4} } \left(2\frac{{ d}x^{i} }{{
d}\mu } \frac{{ d}\omega ^{j} }{{ d}\mu } +\frac{{ d}\omega ^{i}
}{{ d}\mu } \frac{{ d}\omega ^{j} }{{ d}\mu } \right)(\omega ^{1}
)^{\beta _{1} } \ldots (\omega ^{4} )^{\beta _{4} }
\right]\right\}{ d}\mu ,{\kern 1pt} {\kern 1pt} {\kern 1pt} {\kern
1pt}
\end{equation}
where $\mu _{0} ,\mu _{1} $ are values of the affine parameter in
points, which are linked by found geodesic. The sum of terms
containing variations $\omega ^{i} $, $d\omega ^{i} /d\mu $ in
first power is equated to null. Under condition $\omega ^{i} (\mu
_{0} )=\omega ^{i} (\mu _{1} )=0$, this leads to the geodesic
equations in form
\begin{equation} \label{GrindEQ__3_}
\frac{d^{2} x^{l} }{d\mu ^{2} } +\Gamma _{ij}^{l} \frac{dx^{i} }{d\mu } \frac{dx^{j} }{d\mu } =0,
\end{equation}
where $\Gamma _{ij}^{l}$ are Christoffel symbols:
\begin{equation} \label{GrindEQ__4_}
\Gamma _{ij}^{l} =\frac{1}{2} g^{lm} \left(\frac{\partial g_{jm} }{\partial x^{i} } +\frac{\partial g_{im} }{\partial x^{j} } -\frac{\partial g_{ij} }{\partial x^{m} } \right).                                                                                            
\end{equation}

The other terms of series in (\ref{GrindEQ__2_}), containing
variations of coordinates and their derivatives by $\mu $ in more
high powers or their products and being able to have nonzero
values, don't take into account. Thus such method admits violation
of the condition $\eta =0$, which means that with certain
coordinates variations the interval a prior becomes time-like or
space-like. Since this interval accords with the light ray motion,
one leads to the Lorentz-invariance violation in locality, namely,
anisotropies.

It is shown for the massive particle [5] that a fundamental
space-time discreteness need not contradict Lorentz invariance,
and causal set's discreteness is in fact locally Lorentz
invariant. The possibility of Lorentz symmetry break for the
photon in vacuum by effects from the Plank scale is studied in [3,
4]. However, experiments [6] show exceptionally high precision of
constancy of the light speed confirmed a Lorentz symmetry in
locality, and astrophysical tests don't detect isotropic Lorentz
violation [4].

In the method of calculus of variations in the large [7] ones are considered as possible paths along the manifold disregarding kind of interval, not as the trajectories of physical particles. This approach exceeds the limits of classical variational principle in mechanics, according as which virtual motions of the system are compared with cinematically possible motions.

Approximating time-like interval conforming in general relativity to the material particle motion between fixed points to null leads in physical sense to unlimited increase of its momentum, and the space-like interval doesn't conform to move of any object. In this connection it should pay attention on speculation that discreteness at the Planck scale reveals maximum value of momentum for the fundamental particles [8].

The geodesic line must be extremal [1], and the test particle moves along it only in the absence of non-gravity forces. Should photon have some rest mass variations of its path don't give different kinds of intervals, but this assumption doesn't confirm by experiments [9]. We examine choosing of energy so in order that application of variational principle to its integral for deriving of the isotropic critical curves equations would not lead to considering non-null paths.

\section{Definition of Energy and its Variation}

The interval in pseudo-Riemann space-time with metrical coefficients $\tilde{g}_{11} $:
\begin{equation} \label{GrindEQ__5_}
ds^{2} =\tilde{g}_{ij} dx^{i} dx^{j}
\end{equation}
after substitutions
\begin{equation} \label{GrindEQ__6_}
\tilde{g}_{11} =\rho ^{2} g_{11} ,{\kern 1pt} {\kern 1pt} {\kern 1pt} \tilde{g}_{1k} =\rho g_{1k} ,{\kern 1pt} {\kern 1pt} {\kern 1pt} \tilde{g}_{kq} =g_{kq}
\end{equation}
is rewritten in form
\begin{equation} \label{GrindEQ__7_}
ds^{2} =\rho ^{2} g_{11} dx^{12} +2\rho g_{1k} dx^{1} dx^{k} +g_{kq} dx^{k} dx^{q} .
\end{equation}
Here, $\rho $ is some quantity, which is assumed to be equal 1.
Putting down $x^{1} $as time, coordinates with indexes $k,q=2,3,4$
as space coordinates and considering $\rho $ as energy of
light-like particle with $ds=0$ [10,11] we present it as
\begin{equation} \label{GrindEQ__8_}
\rho =\left(g_{11} \frac{dx^{1} }{d\mu } \right)^{-1} \left\{-g_{1k} \frac{dx^{k} }{d\mu } +\right. \left. \sigma \left[(g_{1k} g_{1q} -g_{11} g_{kq} )\frac{dx^{k} }{d\mu } \frac{dx^{q} }{d\mu } \right]^{1/2} \right\}{\kern 1pt} ,
\end{equation}
where $\sigma $ is $\pm 1$.

Indexes except $k,q$ are assigned values 1 to 4. With denotation
of the velocity four-vector components as $u^{i} =dx^{i} /d\mu $
energy variation will be
\begin{equation} \label{GrindEQ__9_}
\delta \rho =\frac{\partial \rho }{\partial x^{\lambda } } \delta x^{\lambda } +\frac{\partial \rho }{\partial u^{\lambda } } \delta u^{\lambda } .
\end{equation}
After substitution
\begin{equation} \label{GrindEQ__10_}
\sigma \left[(g_{1k} g_{1q} -g_{11} g_{kq} )\frac{dx^{k} }{d\mu } \frac{dx^{q} }{d\mu } \right]^{1/2} =g_{1i} \frac{dx^{i} }{d\mu }
\end{equation}
    the partial derivatives with respect to coordinates are written as
\begin{equation} \label{GrindEQ__11_}
\frac{\partial \rho }{\partial x^{\lambda } } =\frac{1}{g_{11} u^{1} } \left[-\frac{\partial g_{1k} }{\partial x^{\lambda } } u^{k} +\frac{1}{2u_{1} } \left(2\frac{\partial g_{1k} }{\partial x^{\lambda } } g_{1q} -\right. \right. \left. \left. \frac{\partial g_{11} }{\partial x^{\lambda } } g_{kq} -\frac{\partial g_{kq} }{\partial x^{\lambda } } g_{11} \right)u^{k} u^{q} \right]-\frac{1}{g_{11} } \frac{\partial g_{11} }{\partial x^{\lambda } } {\kern 1pt} .
\end{equation}
This expression is reduced to
\begin{equation} \label{GrindEQ__12_}
\frac{\partial \rho }{\partial x^{\lambda } } =-\frac{1}{2u_{1} u^{1} } \frac{\partial g_{ij} }{\partial x^{\lambda } } u^{i} u^{j} {\kern 1pt} .
\end{equation}
The partial derivatives with respect to components of the velocity
four-vector are
\begin{equation} \label{GrindEQ__13_}
\frac{\partial \rho }{\partial u^{\lambda } } =-\frac{u_{\lambda } }{u_{1} u^{1} } {\kern 1pt} .
\end{equation}

With $g_{11} =0$ and $g_{1k} \ne 0$ even if for one $k$ the energy takes form
\begin{equation} \label{GrindEQ__14_}
\rho =\frac{g_{kq} u^{k} u^{q} }{2u_{1} u^{1} } .
\end{equation}
In this case the partial derivatives of $\rho $ coincide with
(\ref{GrindEQ__12_}) and (\ref{GrindEQ__13_}).

  For the free moving a particle lagrangian is taken in form
\begin{equation} \label{GrindEQ__14_}
L=-\rho ,
\end{equation}
and conforms to relation [12]:
\begin{equation} \label{GrindEQ__15_}
\rho =u^{\lambda } \frac{\partial L}{\partial u^{\lambda } } -L.
\end{equation}
Thus energy $\rho $ is a hamiltonian of the particle in
gravitational field also an integral of the motion. Obtained
derivatives give the canonical momenta
\begin{equation} \label{GrindEQ__16_}
p_{\lambda } =\frac{\partial L}{\partial u^{\lambda } } =\frac{u_{\lambda } }{u^{1} u_{1} } {\kern 1pt}
\end{equation}
and forces
\begin{equation} \label{GrindEQ__17_}
F_{\lambda } =\frac{\partial L}{\partial x^{\lambda } } =\frac{1}{2u^{1} u_{1} } \frac{\partial g_{ij} }{\partial x^{\lambda } } u^{i} u^{j} {\kern 1pt} .
\end{equation}
We note that the canonical forces, unlike canonical momenta, do not depend on the affine parameter.

Components of the associated vector of the canonical momenta are
\begin{equation} \label{GrindEQ__18_}
p^{\lambda } =\frac{u^{\lambda } }{u^{1} u_{1} } .
\end{equation}

\section{Definition of Momenta and Forces}

Units is chosen so that a light velocity and a Planck constants are $c=h=1$. Physical energy and momenta of photon with frequency $\nu $ in Minkowski space-time with affine parameter $\mu =t$ form contravariant 4-vector of momenta $\pi ^{i} =\nu u^{i} $. For arbitrary affine parameter it is rewritten as
\begin{equation} \label{GrindEQ__19_}
\pi ^{i} =\nu \frac{u^{i} }{u^{1} } .
\end{equation}
These energy and momenta are found from non-gravitational interaction of the particles. We designate them as effective energy and momenta.

    And in pseudo-Riemannian space-time similar energy and momenta of the photon will be put in line with the components of the vector of canonical momenta with raised indices. A certain fixed value of the photon's frequency $\nu _{0} $ is given by the corresponding equality $\nu =\nu _{0} /u_{1} $. Comparing expressions (\ref{GrindEQ__18_}) and (\ref{GrindEQ__19_}), we obtain
\begin{equation} \label{GrindEQ__20_}
\pi ^{i} =\nu _{0} p^{i} .
\end{equation}
This one provides lagrangian of the photon $L_{ph} =\nu _{0} L$.
The components of vector
\begin{equation} \label{GrindEQ__22_}
F^{k} =g^{k\lambda } F_{\lambda }
\end{equation}
associated to (\ref{GrindEQ__17_}), with this approach, are
proportional to gravity forces:
\begin{equation} \label{GrindEQ__23_}
Q_{}^{i} =\nu _{0} F^{i} ,
\end{equation}
which acts on the photon. That is, although non-straight motion of
particle in space-time according to the general relativity due to
its curvature, identified with the gravitational field, we believe
that it is caused by the action of forces obtained by considering
the movement in the coordinate frame.

\section{Equations of Isotropic Critical Curve}

Taking into account equation (\ref{GrindEQ__14_}) a motion
equations are found by using Hamilton's principle from variation
of energy integral
\begin{equation} \label{GrindEQ__24_}
S=\int _{\mu _{0} }^{\mu _{1} } \rho d\mu .
\end{equation}
Energy $\rho $ is non-zero, its variations leave interval to be
light-like, and application of standard variational procedure
yields Euler-Lagrange equations
\begin{equation} \label{GrindEQ__25_}
\frac{d}{d\mu } \frac{\partial \rho }{\partial u^{\lambda } } -\frac{\partial \rho }{\partial x^{\lambda } } =0{\kern 1pt} .
\end{equation}

Critical curve equations are obtained by substitution of partial
derivatives (\ref{GrindEQ__12_}) and (\ref{GrindEQ__13_}) in these
equations. For derivative of the first component of four-velocity
vector we have
\begin{equation} \label{GrindEQ__26_}
\frac{du^{1} }{d\mu } +\frac{u^{1} }{2u_{1} } \frac{\partial g_{ij} }{\partial x^{1} } u^{i} u^{j} =0{\kern 1pt} .
\end{equation}
In the general form, the equations (\ref{GrindEQ__25_}) will be
\begin{equation} \label{GrindEQ__27_}
(g_{1k} v_{\lambda } -g_{k\lambda } v_{1} )\frac{dv^{k} }{d\mu } +\left[\left(\frac{\partial g_{1j} }{\partial x^{i} } -\frac{v_{1} }{2v^{1} } \frac{dv^{1} }{d\mu } \right)v_{\lambda } \right. -\left. \left(\frac{\partial g_{\lambda i} }{\partial x^{j} } -\frac{1}{2} \frac{\partial g_{ij} }{\partial x^{\lambda } } \right)v_{1} \right]v^{i} v^{j} =0.                                     
\end{equation}
Replacement of derivative $du^{1} /d\mu $ here on its expression obtained from (\ref{GrindEQ__26_}) gives
\begin{equation} \label{GrindEQ__28_}
(g_{1k} v_{\lambda } -g_{k\lambda } v_{1} )\frac{dv^{k} }{d\mu } +\left[\left(\frac{\partial g_{1j} }{\partial x^{i} } -\frac{1}{2} \frac{\partial g_{ij} }{\partial x^{1} } \right)v_{\lambda }-\left(\frac{\partial g_{\lambda i} }{\partial x^{j} } -\frac{1}{2} \frac{\partial g_{ij} }{\partial x^{\lambda } } \right)v_{1} \right]v^{i} v^{j} =0.                                       
\end{equation}

The components of these equations can be expressed in terms of the components of the geodetic equations. After multiplication by $g_{kl} $ and summations over repeated indexes $l$ the geodesics equations (\ref{GrindEQ__3_}) will be as follows
\begin{equation} \label{GrindEQ_2_8} 
g_{kj} \frac{d^{2} x^{j} }{d\mu ^{2} } +\frac{1}{2} \left(\frac{\partial g_{ki} }{\partial x^{j} } +\frac{\partial g_{kj} }{\partial x^{i} } -\frac{\partial g_{ji} }{\partial x^{k} } \right)\frac{dx^{j} }{d\mu } \frac{dx^{i} }{d\mu } =0.                                                                        
\end{equation} 
We denote the left-hand side of this equation by $\stackrel{\frown}{D}_{k}.$ Then the equations (\ref{GrindEQ__28_}) are written in the form 
\begin{equation} \label{GrindEQ__29} 
v_{\lambda } \stackrel{\frown}{D}_{1} -v_{1} \stackrel{\frown}{D}_{\lambda } =0.                                                                                                                        
\end{equation}
Coupled with equation (\ref{GrindEQ__26_}), they describe the motion of a light-like particle in accordance with the principle of an extremal energy integral. It can be argued that any solution of null geodesic equations will be such for the obtained equations. According to Cauchy's theorem these systems are identical.

\section{Comparison of Energy Integral Variation and Fermat principles}

Let us clear whether proposed variational method conforms to Fermat's principle for stationary gravity field [1,13], which is formulated as follows
\begin{equation} \label{GrindEQ__29_}
\delta \int  \frac{1}{g_{11} } \left(dl+g_{1k} dx^{k} \right)=0,
\end{equation}
where $dl$ is element of spatial distance along the ray
\begin{equation} \label{GrindEQ__30_}
dl^{2} =\left(\frac{g_{1p} g_{1q} }{g_{11} } -g_{pq} \right)dx^{p} dx^{q} .
\end{equation}
Denoting
\begin{equation} \label{GrindEQ__31_}
df=\frac{1}{g_{11} } \left(dl+g_{1k} dx^{k} \right),
\end{equation}
and comparing this expression with (\ref{GrindEQ__8_}) we write
\begin{equation} \label{GrindEQ__32_}
\frac{df}{d\mu } =-\rho u^{1} .
\end{equation}
Therefore, variation (\ref{GrindEQ__29_}) is equivalent to
variation of integral
\begin{equation} \label{GrindEQ__33_}
S_{1} =\int _{\mu {\kern 1pt} _{0} }^{\mu _{1} } \rho u^{1} d\mu .
\end{equation}

We condition by appropriate choice of the affine parameter $\mu $
the constant value of $u^{1} $. The metrical coefficients in case
of the stationary field doesn't depend on  time, therefore we have
$\partial \rho /\partial x^{1} =0$. The Euler-Lagrange equation
for the $f$ , corresponded to the time coordinate, gives
\begin{equation} \label{GrindEQ__34_}
\frac{d\rho }{d\mu } -\frac{\partial \rho }{\partial x^{1} } u^{1} =0.
\end{equation}
Since the energy $\rho $ is assumed to be constant along critical
curve, its differential is zero. Thus, expression
(\ref{GrindEQ__33_}) with constant $u^{1} $ is idential equation
as well as equation (\ref{GrindEQ__26_}). For the space
coordinates, the equations are follows:
\begin{equation} \label{GrindEQ__35_}
\frac{d}{d\mu } \left(\frac{\partial \rho }{\partial u^{k} } \right)u^{1} +\frac{\partial \rho }{\partial u^{k} } \frac{du^{1} }{d\mu } -\frac{\partial \rho }{\partial x^{k} } u^{1} =0.
\end{equation}
The second term in the left part of equations will be
vanishing and they shall be identical to (\ref{GrindEQ__25_}).

In [14] the generalized Fermat's principle is proposed. It is applied Pontryagin's minimum principle of the optimal control theory and obtained an effective Hamiltonian for the light-like particle motion in a curved spacetime. The dynamical equations for this Hamiltonian are 
\begin{equation} \label{GrindEQ__351}
Q=u^1
\end{equation}
and
\begin{equation}\label{GrindEQ__352}
{d \over d\mu}\left({\partial {Q}\over \partial\dot{x}^q}\right)- {\partial {Q}\over \partial{x}^q}
-{\partial {Q}\over \partial {x}^0}{\partial {Q}\over \partial\dot{x}^q}=0\, .
\end{equation}
Function $Q$ coincides with $-{df}{d\mu }$, under condition that the metric coefficients in(\ref{GrindEQ__30_}), (\ref{GrindEQ__31_}) also depend on time. It is shown that 
obtained dynamical equations correspond to the null geodesic.
Following from (\ref{GrindEQ__32_}) expression for energy $\rho=Q/u^{1}$ substituted in (\ref{GrindEQ__35_}) yields equations (\ref{GrindEQ__352}), which confirms the identity of principle of an extremal energy integral of light-like particle and generalized Fermat's principle.

\section{Energy and Momentum of Particle Transferred to Gravity Field}

 Euler-Lagrange equations can be rewritten in form

\begin{equation} \label{GrindEQ__39_}
\frac{dp_{\lambda } }{d\mu } -F_{\lambda } =0\, \, .
\end{equation}
Passing in these equations to the associated canonical momenta and forces, we obtain
\begin{equation} \label{GrindEQ__40_}
F^{k} =\frac{dp^{k} }{d\mu } +g^{k{\kern 1pt} \lambda } \frac{dg_{\lambda {\kern 1pt} i} }{d\mu } p^{i} .
\end{equation}

In accordance with conservation laws, the vector of energy and
momentum of a system that includes a particle and the
gravitational field generated by it, denoted by $\bar{p}^{k} $,
can be written as the sum of the momentum and energy of the
particle itself $p^{k} $and transmitted it to the gravitational
field $\stackrel{\leftrightarrow}{p}^{k} $. The vector
$\bar{p}^{k} $ changes under the influence of the force from the
source of gravity:
\begin{equation} \label{GrindEQ__41_}
\frac{d\bar{p}^{k} }{d\mu } =\frac{dp^{k} }{d\mu } +\frac{d\stackrel{\leftrightarrow}{p}^{k} }{d\mu } =F^{k} .
\end{equation}
Comparing two expressions for $F^{k} $ and passing in
(\ref{GrindEQ__40_}) to the partial derivatives of metrical
coefficients we find the rate of exchange of energy and momentum
between particle and gravitational field
\begin{equation} \label{GrindEQ__42_}
\frac{d\stackrel{\leftrightarrow}{p}^{k} }{d\mu } =g^{k{\kern 1pt} \lambda } \frac{\partial g_{\lambda {\kern 1pt} i} }{\partial x^{j} } u^{j} p^{i} .
\end{equation}

   From the conservation laws it follows that the force acting on the system including the particle and the gravitational field generated by it is equal in magnitude and opposite in sign to the force acting on the system of the source of gravitation from the side of the particle system. This is equivalent to fulfilling Newton's third law. Its adherence to the Newtonian limit of gravity means the equality of the passive and active gravitational masses.

\section{Photon's Dynamics in Schwarzschild Space-Time}

\subsection{ Spherical Coordinates}

A centrally symmetric gravity field in the free space is described by the Schwarzschild metric. At spherical coordinates $x^{i} =(t,r,\theta ,\varphi )$ its line element is
\begin{equation} \label{GrindEQ__43_}
ds^{2} =\left(1-\frac{\alpha }{r} \right)dt^{2} -\left(1-\frac{\alpha }{r} \right)^{-1} dr^{2} -r^{2} (d\theta ^{2} +\sin ^{2} \theta d\phi ^{2} ),
\end{equation}
where $\alpha $ is constant.

For this space we find equations of the critical curve of the
integral energy $\rho $. The canonical momenta
(\ref{GrindEQ__16_}) for the cyclic coordinates $t,\varphi $ are
the constants of motion
\begin{equation} \label{GrindEQ__44_}
A=\frac{dt}{d\mu } {\kern 1pt} ,
\end{equation}
\begin{equation} \label{GrindEQ__45_}
C=r^{2} \sin ^{2} \theta \frac{d\varphi }{d\mu } \left(1-\frac{\alpha }{r} \right)^{-1} {\kern 1pt} .
\end{equation}
Equations (\ref{GrindEQ__28_}) for coordinates $r,\theta $ give
\begin{equation} \label{GrindEQ__46_}
\frac{d^{2} r}{d\mu ^{2} } +\frac{\alpha }{2r^{2} } \left(1-\frac{\alpha }{r} \right)\left(\frac{dt}{d\mu } \right)^{2} -\left. \frac{3\alpha }{2r(r-\alpha )} \left(\frac{dr}{d\mu } \right)^{2} -(r-\alpha )\left[\left(\frac{d\theta }{d\mu } \right)^{2} +\right. \sin ^{2} \theta \left(\frac{d\phi }{d\mu } \right)^{2} \right]=0{\kern 1pt} ,
\end{equation}
\begin{equation} \label{GrindEQ__47_}
\frac{d^{2} \theta }{d\mu ^{2} } +\frac{2r-3\alpha }{r(r-\alpha )} \frac{dr}{d\mu } \frac{d\theta }{d\mu } +\frac{1}{2} \sin 2\theta \left(\frac{d\varphi }{d\mu } \right)^{2} =0{\kern 1pt} ,
\end{equation}
Metric (\ref{GrindEQ__43_}) for the isotropic curve yields
\begin{equation} \label{GrindEQ__48_}
\left(1-\frac{\alpha }{r} \right)\left(\frac{dt}{d\mu } \right)^{2} -\left(1-\frac{\alpha }{r} \right)^{-1} \left(\frac{dr}{d\mu } \right)^{2} -r^{2} \left[\left(\frac{d\theta }{d\mu } \right)^{2} +\sin ^{2} \theta \left(\frac{d\phi }{d\mu } \right)^{2} \right]=0{\kern 1pt} .
\end{equation}
Assuming that $A=1$ and considering motion in plane $\theta =\pi
/2$ we write derivatives of the cyclic coordinates
\begin{equation} \label{GrindEQ__49_}
\frac{dt}{d\mu } =1{\kern 1pt} ,
\end{equation}
\begin{equation} \label{GrindEQ__50_}
\frac{d\varphi }{d\mu } =\frac{C}{r^{2} } \left(1-\frac{\alpha }{r} \right){\kern 1pt} .
\end{equation}
Substituting these values in equation (\ref{GrindEQ__48_}) we find
\begin{equation} \label{GrindEQ__51_}
\frac{dr}{d\mu } =\pm \left[\left(1-\frac{\alpha }{r} \right)^{2} -\left(\frac{C}{r} \right)^{2} \left(1-\frac{\alpha }{r} \right)^{3} \right]^{1/2} {\kern 1pt} .
\end{equation}

Found velocities coincide with solutions of the null geodesic equations for the Schwarzschild space-time [2] to within parameter of differentiation
\begin{equation} \label{GrindEQ__52_}
d\mu =d\mu _{s} \left(1-\frac{\alpha }{r} \right)^{-1} ,
\end{equation}
where $\mu _{s} $ corresponds to geodesic equations.

The canonical momenta (\ref{GrindEQ__16_}) and forces
(\ref{GrindEQ__17_}) are
\[p_{1} =1,{\kern 1pt} {\kern 1pt} {\kern 1pt} {\kern 1pt} p_{2} =\mp \frac{1}{\left(1-\frac{\alpha }{r} \right)} \sqrt{1-\frac{C^{2} }{r^{2} } \left(1-\frac{\alpha }{r} \right)} ,\]
\begin{equation} \label{GrindEQ__53_}
p_{3} =0,{\kern 1pt} {\kern 1pt} {\kern 1pt} {\kern 1pt} p_{4} =-C;
\end{equation}
\[F_{1} =F_{3} =F_{4} =0,\]
\begin{equation} \label{GrindEQ__54_}
F_{2} =\frac{\alpha }{r^{2} \left(1-\frac{\alpha }{r} \right)} -\frac{C^{2} }{r^{3} } +\frac{\alpha {\kern 1pt} C^{2} }{2r^{4} } .
\end{equation}
Nonzero components of vector of the canonical momenta with raised
indices are
\[p^{1} =\frac{1}{\left(1-\frac{\alpha }{r} \right)} ,\]
\[p^{2} =\pm \sqrt{1-\frac{C^{2} }{r^{2} } \left(1-\frac{\alpha }{r} \right)} ,\]
\begin{equation} \label{GrindEQ__55_}
p^{4} =\frac{C}{r^{2} } .
\end{equation}
 A nonzero component of vector of the canonical forces with raised indices is
\begin{equation} \label{GrindEQ__56_}
F^{2} =-\frac{\alpha }{r^{2} } +\frac{C^{2} }{r^{3} } \left(1-\frac{\alpha }{r} \right)\left(1-\frac{\alpha }{2r} \right).
\end{equation}

In so far as with gravitational constant $G$ and active
gravitational mass $M$ the Newtonian limit of gravity theory
requires $\alpha =2GM$, for the radial motion, $C=0$, the first
term of $F^{2} $ yields twice Newton gravity force. Taking into
account (\ref{GrindEQ__23_}) it corresponds to the gravitatinal
mass of the photon
\begin{equation} \label{GrindEQ__57_}
m_{g{\kern 1pt} p}^{} =2\nu _{0} .
\end{equation}

\subsection{ Rectangular Coordinates}

 Considering the non-radial motion in order to avoid the appearance of a fictitious component of the force due to the sphericity of the coordinate system, we use the Schwarzschild metric in rectangular coordinates. The isotropic form of the Schwarzschild metric, to which one can go from its standard form (\ref{GrindEQ__43_}) with the help of the transformation
\begin{equation} \label{GrindEQ__58_}
r=\left(1+\frac{\alpha }{4\bar{r}} \right)^{2} \bar{r},
\end{equation}
is written as
\begin{equation} \label{GrindEQ__59_}
ds^{2} =\left(\frac{1-\frac{\alpha }{4\bar{r}} }{1+\frac{\alpha }{4\bar{r}} } \right)^{2} dt^{2} -\left(1+\frac{\alpha }{4\bar{r}} \right)^{4} (dx^{2} +dy^{2} +\, {\kern 1pt} dz^{2} ),
\end{equation}
where $(t,x,y,z)$ is rectangular frame and $\bar{r}=\sqrt{x^{2} +y^{2} +z^{2} } $.

 We will consider the motion in the plane $z=0$ and seek the force acting on the particle at a point $(t,x,0,0)$ that corresponds to the value of the angular coordinate $\varphi =0$ in the spherical frame. Coordinate transformations in the plane are
\begin{equation} \label{GrindEQ__60_}
x=\bar{r}\cos \varphi ,\quad \quad y=\bar{r}\sin \varphi .
\end{equation}
The nonzero spatial components of the 4-velocity are
\begin{equation} \label{GrindEQ__61_}
\bar{u}^{2} =\frac{dx}{d\mu } =\frac{d\bar{r}}{d\mu } ,\quad \quad \bar{u}^{3} =\frac{dy}{d\mu } =\frac{d\varphi }{d\mu } \bar{r}.
\end{equation}
The transformation (\ref{GrindEQ__58_}) implies the relation
\begin{equation} \label{GrindEQ__62_}
dr=\left(1-\frac{\alpha ^{2} }{16\bar{r}^{2} } \right)d\bar{r}.
\end{equation}
Equations (\ref{GrindEQ__49_})-(\ref{GrindEQ__51_}) yield
\begin{equation} \label{GrindEQ__63_}
\bar{u}^{1} =1,
\end{equation}
\begin{equation} \label{GrindEQ__64_}
\bar{u}_{1} =\left(\frac{1-\frac{\alpha }{4\bar{r}} }{1+\frac{\alpha }{4\bar{r}} } \right)^{2}  ,
\end{equation}
\begin{equation} \label{GrindEQ__65_}
\bar{u}^{2} =\pm \frac{\left(1-\frac{\alpha }{4\bar{r}} \right)}{\left(1+\frac{\alpha }{4\bar{r}} \right)^{3} } \left[1\, -\frac{C^{2} \left(1-\frac{\alpha }{4\bar{r}} \right)^{2} }{\bar{r}^{2} \left(1+\frac{\alpha }{4\bar{r}} \right)^{6} } \right]^{1/2} ,
\end{equation}
\begin{equation} \label{GrindEQ__66_}
\bar{u}^{3} =\frac{C\left(1-\frac{\alpha }{4\bar{r}} \right)^{2} }{\bar{r}\left(1+\frac{\alpha }{4\bar{r}} \right)^{6} }  .
\end{equation}

Substitution of these velocities in (\ref{GrindEQ__18_}) gives components of associated vector of the canonical momenta 
\begin{equation} \label{GrindEQ__58_} 
{\overline{p}}^{\mathrm{1}}\mathrm{=}{\left(\frac{\mathrm{1+}\frac{\alpha }{\mathrm{4}\overline{r}}}{\mathrm{1-}\frac{\alpha }{\mathrm{4}\overline{r}}}\right)}^{\mathrm{2}}, 
\end{equation} 
\begin{equation} \label{GrindEQ__59_} 
{\overline{p}}^{\mathrm{2}}\mathrm{=\pm }\frac{\mathrm{1}}{\left(\mathrm{1-}\frac{\alpha }{\mathrm{16}{\overline{r}}^{\mathrm{2}}}\right)}{\left[\mathrm{1\ -}\frac{C^{\mathrm{2}}{\left(\mathrm{1-}\frac{\alpha }{\mathrm{4}\overline{r}}\right)}^{\mathrm{2}}}{{\overline{r}}^{\mathrm{2}}{\left(\mathrm{1+}\frac{\alpha }{\mathrm{4}\overline{r}}\right)}^{\mathrm{6}}}\right]}^{\mathrm{1/2}}, 
\end{equation} 
\begin{equation} \label{GrindEQ__60_} 
{\overline{p}}^{\mathrm{3}}\mathrm{=}\frac{C}{\overline{r}{\left(\mathrm{1+}\frac{\alpha }{\mathrm{4}\overline{r}}\right)}^4}. 
\end{equation} 
Passing back from the variable $\overline{r}$ to $r$, we write, in accordance with equation (\ref{GrindEQ__20_}), the value of the photon energy and momentum in a remote coordinate frame 
\begin{equation} \label{GrindEQ__61_} 
E\mathrm{=}{\nu }_0{\left(\mathrm{1-}\frac{\alpha }{r}\right)}^{\mathrm{-}\mathrm{1}}, 
\end{equation} 
\begin{equation} \label{GrindEQ__62_} 
\overline{P}\mathrm{=}{\mathrm{[(}{\overline{p}}^{\mathrm{2}}\mathrm{)+}{(\overline{p}}^{\mathrm{3}}\mathrm{)]}}^{\mathrm{1/2}}\mathrm{=}\frac{\nu}{\left(\mathrm{1-}\frac{\alpha }{\mathrm{16}{\overline{r}}^{\mathrm{2}}}\right)}, 
\end{equation} 
 where ${\nu }_0$ is the photon frequency at infinity at the world line with unlimited $r$. Moving to the scale of the length of spherical frame in view of Eq. (\ref{GrindEQ__62_}) we obtain $P=\nu _{0}$.

The components of the canonical forces vector $F^{k} $ (\ref{GrindEQ__22_}) are put in correspondence with the gravitational forces acting on the particle. Substituting these 4-velocities in (\ref{GrindEQ__17_}), we find the unique nonzero component of the force vector acting on the light-like particle:
\begin{equation} \label{GrindEQ__67_}
\bar{F}_{}^{2} =-\frac{\alpha \left(1-\frac{\alpha }{8\bar{r}} \right)}{\bar{r}^{2} \left(1+\frac{\alpha }{4\bar{r}} \right)^{5} \left(1-\frac{\alpha }{4\bar{r}} \right)} .
\end{equation}
Its magnitude does not depend on the direction of motion of the
photon. This formula differs from equation (\ref{GrindEQ__56_})
with $C=0$, which corresponds to the radial motion in spherical
coordinates. That is, the expression for the gravitational force
acting on the photon depends on the choice of the coordinate
system. However, in the limit of weak gravity these expressions
asymptotically converge and give Newton's law of gravitation with
the gravitational mass of the photon (\ref{GrindEQ__57_}). One
conforms to the light deflection in central gravity field [15],
which is twice value being given by the Newton gravity theory.

  Obtained gravitational mass of the light-like particle is independent on the direction of its motion. The gravitational mass of a photon for low gravity is equal to doubled mass of a material particle, equivalent to its energy. This corresponds to the results of Tolman [16] obtained for the interaction between a light package and a material particle.

\section{ Extremal Isotropic Curves in FLRW
Space-Time}

The FLRW cosmological model for the flat space with rectangular coordinates $x^{i} =(t,x^{q} )$ is described by metric
\begin{equation} \label{GrindEQ__68_}
ds^{2} =dt^{2} -a^{2} (t)dx^{q2} ,
\end{equation}
where $a$ is the length scale factor.

Equation of motion of the light-like particle for the time
coortinate (\ref{GrindEQ__26_}) gives
\begin{equation} \label{GrindEQ__69_}
\frac{d^{2} t}{d\mu ^{2} } -\dot{a}a\left(\frac{dx^{q} }{d\mu } \right)^{2} =0,
\end{equation}
where overdot denotes derivative with respect to time. The
Euler-Lagrange equations (\ref{GrindEQ__39_}) for the cyclic
coordinates $x^{q} $ yield constants of motion
\begin{equation} \label{GrindEQ__70_}
p_{q} =-a^{2} \frac{dx^{q} }{d\mu } /\left(\frac{dt}{d\mu } \right)^{2} .
\end{equation}
Having extracted derivatives with respect to the space-like
coordinates from this equation and substituting them in
(\ref{GrindEQ__69_}) we obtain
\begin{equation} \label{GrindEQ__71_}
\frac{d^{2} t}{d\mu ^{2} } -p_{q}^{2} \frac{\dot{a}}{a^{3} } \left(\frac{dt}{d\mu } \right)^{4} =0.
\end{equation}
This equation has solution, which with denotation $\Pi =p_{q}^{2}
$ is written in form
\begin{equation} \label{GrindEQ__72_}
\frac{dt}{d\mu } =\left(\Pi a^{-2} +B\right)^{-1/2} ,
\end{equation}
where $B$ is constant. Substitution found first component of the
four-velocity vector in equation (\ref{GrindEQ__70_}) gives
\begin{equation} \label{GrindEQ__73_}
\frac{dx^{q} }{d\mu } =-p_{q} a^{-2} \left(\Pi a^{-2} +B\right)^{-1} .
\end{equation}
The condition, following from Eq. (\ref{GrindEQ__68_}):
\begin{equation} \label{GrindEQ__74_}
p_{q} =\left(\frac{dt}{d\mu } \right)^{2} -a^{2} \left(\frac{dx^{q} }{d\mu } \right)^{2} ,
\end{equation}
corresponds to isotropic curve. It yields $B=0$ and components of
the four-velocity vector turn out to
\begin{equation} \label{GrindEQ__75_}
\frac{dt}{d\mu } =\frac{1}{\Pi ^{1/2} } a,
\end{equation}
\begin{equation} \label{GrindEQ__76_}
\frac{dx^{q} }{d\mu } =-\frac{p_{q} }{\Pi } .
\end{equation}
They conform to solution of equations of the null geodesics for
the FLRW space-time [2].

The canonical momenta of the light-like particle are
\begin{equation} \label{GrindEQ__77_}
p_{1} =\Pi ^{1/2} a^{-1}
\end{equation}
and constant $p_{q} $. The canonical forces are
\begin{equation} \label{GrindEQ__78_}
F_{1} =\frac{\dot{a}}{a} \Pi ,{\kern 1pt} {\kern 1pt} {\kern 1pt} {\kern 1pt} F_{q} =0.
\end{equation}
Their associated values is written as
\begin{equation} \label{GrindEQ__79_}
p^{1} =\Pi ^{1/2} a^{-1} ,{\kern 1pt} {\kern 1pt} {\kern 1pt} {\kern 1pt} p^{q} =-p_{q} a^{-2}
\end{equation}
and
\begin{equation} \label{GrindEQ__80_}
F^{1} =\frac{\dot{a}}{a} \Pi ,{\kern 1pt} {\kern 1pt} {\kern 1pt} {\kern 1pt} F^{q} =0.
\end{equation}

\section{Extremal Isotropic Curves in Goedel Space-Time}

The stationary solution of the Einstein's field equation with
cosmological constant found by Goedel describes gravity field of
the rotating uniform dust matter. With coordinates $x^{i}
=(t,r,y,z)$ the line element is written in form
\begin{equation} \label{GrindEQ__81_}
ds^{2} =dt^{2} -dr^{2} -dz^{2} +2\exp(\sqrt{2} \omega r) dtdy+\frac{1}{2} \exp(2\sqrt{2} \omega r) dy^{2} ,
\end{equation}
where $\omega $ is constant.

\subsection{Solution by Use of Principle of Stationary Integral of Energy}

The canonical momenta (\ref{GrindEQ__16_}) for cyclic coordinates
$t,y,z$ are the constants of motion. They are written in form
\[p_{1} =\frac{1}{u^{1} } ,\]
\[p_{3} =\frac{\exp(\sqrt{2} \omega r) u^{1} +\frac{1}{2} \exp(2\sqrt{2} \omega r) u^{3} }{u^{1} \left(u^{1} +\exp(\sqrt{2} \omega r) u^{3} \right)} ,\]
\begin{equation} \label{GrindEQ__82_}
p_{4} =-\frac{u^{4} }{u^{1} \left(u^{1} +\exp(\sqrt{2} \omega r) u^{3} \right)} .
\end{equation}
These equations with following from Eq. (\ref{GrindEQ__81_})
condition
\begin{equation} \label{GrindEQ__83_}
0=(u^{1} )^{2} -(u^{2} )^{2} -(u^{4} )^{2} +2\exp(\sqrt{2} \omega r) u^{1} u^{3} +\frac{1}{2} \exp(2\sqrt{2} \omega r) (u^{3} )^{2}
\end{equation}
yield components of the four-velocity vector:
\begin{equation} \label{GrindEQ__84_}
\frac{dt}{d\mu } =\frac{1}{p_{1} } ,
\end{equation}
\begin{equation} \label{GrindEQ__85_}
\frac{dr}{d\mu } ==\pm \frac{\left[4p_{1} p_{3} \exp(\sqrt{2} \omega r) -(p_{1}^{2} +p_{4}^{2} )\exp(2\sqrt{2} \omega r) -2p_{3}^{2} \right]^{1/2} }{p_{1} \left(p_{1} \exp(\sqrt{2} \omega r) -2p_{3} \right)} ,
\end{equation}
\begin{equation} \label{GrindEQ__86_}
\frac{dy}{d\mu } =2\frac{p_{3} -p_{1} \exp(\sqrt{2} \omega r) }{p_{1} \exp(\sqrt{2} \omega r) \left(p_{1} \exp(\sqrt{2} \omega r) -2p_{3} \right)} ,
\end{equation}
\begin{equation} \label{GrindEQ__87_}
\frac{dz}{d\mu } =\frac{p_{4} \exp(\sqrt{2} \omega r) }{p_{1} \left(p_{1} \exp(\sqrt{2} \omega r) -2p_{3} \right)} .
\end{equation}
With $p_{1} \exp(\sqrt{2} \omega r) =2p_{3} $ the singularity takes
place.

  The canonical momentum corresponding to coordinate $r$ is
\begin{equation} \label{GrindEQ__88_}
p_{2} =\pm \left[4p_{1} p_{3} \exp(-\sqrt{2} \omega r) -(p_{1}^{2} +p_{4}^{2} )-\right. \left. 2p_{3}^{2} \exp(-2\sqrt{2} \omega r) \right]^{1/2} .
\end{equation}
Canonical forces have values
\[F_{1} =F_{3} =F_{4} =0,\]
\begin{equation} \label{GrindEQ__89_}
F_{2} =2\sqrt{2} \omega \frac{p_{3} \left(p_{3} -p_{1} \exp(\sqrt{2} \omega r) \right)}{\left(p_{1} \exp(\sqrt{2} \omega r) -2p_{3} \right)^{2} } .
\end{equation}
Associated canonical momentum and forces are
\[p^{1} =-p_{1} +2p_{3} \exp(-\sqrt{2} \omega r) ,\]
\[p^{2} =\mp \left[4p_{1} p_{3} \exp(-\sqrt{2} \omega r) -(p_{1}^{2} +p_{4}^{2} )-\right. \left. 2p_{3}^{2} \exp(-2\sqrt{2} \omega r) \right]^{1/2} ,\]
\[p^{3} =2p_{1} \exp(-\sqrt{2} \omega r) -2p_{3} \exp(-2\sqrt{2} \omega r) ,\]
\begin{equation} \label{GrindEQ__90_}
p^{4} =-p_{4} ;
\end{equation}
\[F^{1} =F^{3} =F^{4} =0,\]
\begin{equation} \label{GrindEQ__91_}
F^{2} =-2\sqrt{2} \omega \frac{p_{3} \left(p_{3} -p_{1} \exp(\sqrt{2} \omega r) \right)}{\left(p_{1} \exp(\sqrt{2} \omega r) -2p_{3} \right)^{2} } .
\end{equation}

\subsection{Comparision of Extreme Integral of Energy Curves and Geodesics}

 The procedure for obtaining the geodesic equations by the variation of the integral of expression (\ref{GrindEQ__1_}) is identical to finding the Euler-Lagrange equations for the Lagrangian
\begin{equation} \label{GrindEQ__92_}
L_{g} =\eta ,
\end{equation}
that is, these equations are identical. For metric
(\ref{GrindEQ__81_}) we have
\begin{equation} \label{GrindEQ__93_}
L_{g} =\left(\tilde{u}^{1} \right)^{2} -\left(\tilde{u}^{2} \right)^{2} -\left(\tilde{u}^{4} \right)^{2} +2\exp (\sqrt{2} \omega {\kern 1pt} r)\tilde{u}^{1} \tilde{u}^{3} +\frac{1}{2} \exp (2\sqrt{2} \omega {\kern 1pt} r)\left(\tilde{u}^{3} \right)^{2} ,
\end{equation}
where $\tilde{u}^{i} $ are 4-velocities of geodesics. The
constants of motion are
\[\tilde{p}_{1} =\tilde{u}^{1} +\exp (\sqrt{2} \omega r)\tilde{u}^{3} , \]
\[\tilde{p}_{3} =\exp (\sqrt{2} \omega {\kern 1pt} r)\tilde{u}^{1} +\frac{1}{2} \exp (2\sqrt{2} \omega {\kern 1pt} r)\tilde{u}^{3} ,\]
\begin{equation} \label{GrindEQ__94_}
\tilde{p}_{4} =-\tilde{u}^{4} .
\end{equation}
These equations, together with condition (\ref{GrindEQ__83_}) for
4-velocities, yield
\begin{equation} \label{GrindEQ__95_}
\frac{cdt}{d\tilde{\mu }} =-\tilde{p}_{1} +2\tilde{p}_{3} \exp (-\sqrt{2} \omega {\kern 1pt} r),
\end{equation}
\begin{equation} \label{GrindEQ__96_}
\frac{dr}{d\tilde{\mu }} =\pm \left[-\tilde{p}_{1} ^{2} -\tilde{p}_{4} ^{2}+4\tilde{p}_{1} \tilde{p}_{3} \exp (-\sqrt{2} \omega {\kern 1pt} r)-2\tilde{p}_{3} ^{2} \exp (-2\sqrt{2} \omega {\kern 1pt} r)\right]^{1/2} ,
\end{equation}
\begin{equation} \label{GrindEQ__97_}
\frac{dy}{d\tilde{\mu }} =2\left[\tilde{p}_{1} \exp (-\sqrt{2} \omega {\kern 1pt} r)-\tilde{p}_{3} \exp (-2\sqrt{2} \omega {\kern 1pt} r)\right],
\end{equation}
\begin{equation} \label{GrindEQ__98_}
\frac{dz}{d\tilde{\mu }} =-\tilde{p}_{4} .
\end{equation}
As a result, we obtain velocities as the derivatives of spatial coordinates with respect to time
\begin{equation} \label{GrindEQ__99_}
\dot{r}_{g} =\pm \frac{\left[-(\tilde{p}_{1} ^{2} +\tilde{p}_{4} ^{2} )\exp (2\sqrt{2} \omega {\kern 1pt} r)+4\tilde{p}_{1} \tilde{p}_{3} \exp (\sqrt{2} \omega {\kern 1pt} r)-2\tilde{p}_{3} ^{2} \right]^{1/2} }{\tilde{p}_{1} \exp (\sqrt{2} \omega {\kern 1pt} r)-2\tilde{p}_{3} } ,
\end{equation}
\begin{equation} \label{GrindEQ__100_}
\dot{y}_{g} =2\frac{\tilde{p}_{3} -\tilde{p}_{1} \exp (\sqrt{2} \omega {\kern 1pt} r)}{\exp (\sqrt{2} \omega {\kern 1pt} r)\left[\tilde{p}_{1} \exp (\sqrt{2} \omega {\kern 1pt} r)-2\tilde{p}_{3} \right]}  ,
\end{equation}
\begin{equation} \label{GrindEQ__101_}
\dot{z}_{g} =\frac{\tilde{p}_{4} \exp (\sqrt{2} \omega {\kern 1pt} r)}{\tilde{p}_{1} \exp (\sqrt{2} \omega {\kern 1pt} r)-2\tilde{p}_{3} } .
\end{equation}

Solution by use of principle of stationary integral of energy
(\ref{GrindEQ__84_})-(\ref{GrindEQ__87_}) gives velocities
\begin{equation} \label{GrindEQ__102_}
\dot{r}=\pm \frac{\left[-(p_{1} ^{2} +p_{4} ^{2} )\exp (2\sqrt{2} \omega {\kern 1pt} r)+4p_{1} p_{3} \exp (\sqrt{2} \omega {\kern 1pt} r)-2p_{3} ^{2} \right]^{1/2} }{p_{1} \exp (\sqrt{2} \omega {\kern 1pt} r)-2p_{3} } ,
\end{equation}
\begin{equation} \label{GrindEQ__103_}
\dot{y}=2\frac{p_{3} -p_{1} \exp (\sqrt{2} \omega {\kern 1pt} r)}{\exp (\sqrt{2} \omega {\kern 1pt} r)\left[p_{1} \exp (\sqrt{2} \omega {\kern 1pt} r)-2p_{3} \right]} ,
\end{equation}
\begin{equation} \label{GrindEQ__104_}
\dot{z}=\frac{p_{4} \exp (\sqrt{2} \omega {\kern 1pt} r)}{p_{1} \exp (\sqrt{2} \omega {\kern 1pt} r)-2p_{3} } .
\end{equation}
Expressions (\ref{GrindEQ__100_}) and (\ref{GrindEQ__103_}),
(\ref{GrindEQ__99_})-(\ref{GrindEQ__101_}) and (\ref{GrindEQ__102_})-(\ref{GrindEQ__102_})
are identical. Thus, curves of extreme integral of a light-like particle energy coincide with isotropic geodesics in Goedel space-time.

\section{Conclusions}

The proposed form of energy of the light-like particle allows applying of the Lagrange's mechanics for analysis of its motion in the curvilinear space-time. Considered procedure of production of the free motion equations by variation of the energy integral conforms to principles of the calculus of variations in the classic mechanics in accordance with which the motion variations must be cinematically admissible for the system. The virtual displacements of coordinates retain path of the light-like particle to be null in the pseudo-Riemann space-time, i.e. not lead to the Lorentz-invariance violation in locality. Equations, obtained by this method, agrees with result given by the generalized Fermat's and geodesics principles.

   A definite Lagrangian produces particle canonical momenta and forces acting on it in the coordinate frame. Contravariant momenta are identified with effective, determined from the non-gravitational interactions, energy, and momentum of particle in gravitational field, and contravariant forces are mapped to the components of the vector of the gravitational force. Moving in the gravitational field, the particle exchanges energy and momentum with it. The corresponding energy-momentum vector of particle plays the same role as the pseudotensor used in the laws of conservation in tensor form. The value of the force acting on a particle in the coordinate frame depends on the choice of the reference frame, and therefore the quantities determined through them are meaningful only for weak gravity, for which its values asymptotically converge in different coordinate frames. The analogy between the mechanics of particle motion in the Schwarzschild space and Newton's gravity theory allows to determine the gravitational mass of the photon, which is equal to twice the mass of a material particles of the same effective energy.

\begin{acknowledgements}

I acknowledge J. Foukzon, V. B. Morozov, A. Shatalov  for useful discussion.

\end{acknowledgements}


\begin{enumerate}
\item \textbf{ }L.D. Landau and E.M. Lifshitz, Classical Theory of Fields, Fourth Revised English Edition, Oxford: Pergamon, 1975.

\item C.W. Misner, K.S. Thorne  and J.A. Wheeler J A  Gravitation, vol 1, San-Francisco: Freeman and Company, 1973.

\item  A. Kostelecky and A. Pickering, Vacuum Photon Splitting in Lorentz-Violating Quantum Electrodynamics, Phys. Rev. Lett. 91 (2003) 031801, hep-ph/0212382; T. Jacobson, S. Liberati, D. Mattingly, F.W. Stecker, New limits on Planck scale Lorentz violation in QED, Phys. Rev. Lett. 93 (2004) 021101, astro-ph/0309681.

\item  A. Kostelecky, M. Mewes, Astrophysical Tests of Lorentz and CPT Violation with Photons, Astrophys.J.689 (2008) L1-L4, arXiv:0809.2846.

\item  F. Dowker, J. Henson, R. D. Sorkin, Quantum Gravity Phenomenology, Lorentz Invariance and Discreteness, Mod. Phys. Lett. A19 (2004) 1829-1840, gr-qc/0311055; L. Bombelli, J. Henson, R. D. Sorkin, Discreteness without symmetry breaking: a theorem, gr-qc/0605006.

\item   H. Mueller et al. Relativity tests by complementary rotating Michelson-Morley experiments, Phys.Rev.Lett. 99 (2007) 050401, arXiv:0706.2031; P. L. Stanwix et al. Improved test of Lorentz Invariance in Electrodynamics using Rotating Cryogenic Sapphire Oscillators, Phys.Rev. D74 (2006) 081101, gr-qc/0609072; Antonini et al. Test of constancy of speed of light with rotating cryogenic optical resonators, Phys. Rev. A. 71 (2005) 050101.

\item  M. Morse, The Calculus of Variations in the Large, Colloquium Publications of the American Mathematical Society, vol. 18. placeStateNew York, 1934.

\item  G. Amelino-Camelia, Doubly Special Relativity, Nature 418 (2002) 34-35, gr-qc/0207049; J. Magueijo and L. Smolin, Lorentz invariance with an invariant energy scale, Phys. Rev. Lett. 88 (2002) 190403, hep-th/0112090.

\item  A.S. Goldhaber and M.M. Nieto, Photon and Graviton Mass Limits, ArXiv:0809.1003.

\item  W.B. Belayev, Variation of the light-like particle energy and its critical curve equations, arXiv:0806.3350.

\item  W.B. Belayev, {Dinamika v obschei teorii otnositel'nosti: variacionn'ie metod'i} (Moscow: URSS), 2017, (in Russian), [{The dynamics in general relativity theory: variational methods}].

\item  L. D. Landau and E.M. Lifshitz, Vol. 1. Mechanics, 3ed., Oxford: Pergamon, 1976.

\item V. Perlick, Gravitational Lensing from a Spacetime Perspective, {Living Rev. Relativity} 7 (2004), arXiv:1010.3416.

\item V.P. Frolov, Generalized Fermat?s principle and action for light rays in a curved spacetimePhys. Rev. D 88, 064039 (2013),
arXiv:1307.3291.

\item  A. Einstein, Die Grundlage der allgemainen {\rm Relativit$\ddot{a}$tstheorie,} Ann. Phys. 49 (1916) 769-822.

\item  Tolman R.C. Relativity Thermodynamics and Cosmology, Oxford: At the Claredon Press, 1969
\end{enumerate}

\end{document}